\documentstyle[12pt,twoside, a4]{article}
\newcommand{\br} {{\bf r}}

\newcommand{\bp} {{\bf p}}
\newcommand{\by} {{\bf y}}
\newcommand{\bk} {{\bf k}}
\newcommand{\bsigma} {{\mbox{\boldmath $\sigma$}}}

\newcommand{\bnabla} {{\mbox{\boldmath $\nabla$}}}

\newcommand{\be} {{\bf e}}

\newcommand{\ds}{D^{\ast}}
\newcommand{\dss}{D^{\ast \ast}}
\newcommand{\D}{D_{s}}
\newcommand{\Ds}{D^{\ast}_{s}}

\newcommand{\eqntimes}{\mbox{} \times}
\newcommand{\intereqnvspace}{\vspace{-1.4ex}}
\newcommand{\eqnhspace}{\hspace{3em}}
\newcommand{\eqnneghspace}{\hspace{-6em}}
\newcommand{\beqn}{\begin{equation}}
\newcommand{\eeqn}{\end{equation}}
\newcommand{\barr}[1]{\begin{array}{#1}}
\newcommand{\earr}{\end{array}}
\newcommand{\beqna}{\begin{eqnarray}}
\newcommand{\eeqna}{\end{eqnarray}}
\newcommand{\btablec}{\begin{table}[tbp] \begin{center}}
\newcommand{\etablec}{\end{center} \end{table}}

\newcommand{\lapprox}{\stackrel{<}{\scriptstyle \sim}}

\begin{document}

\title{\small \rm \begin{flushright} \small{hep-ph/9502204}\\
\small{OUTP-94-34P} \end{flushright} \vspace{2cm}
\LARGE \bf Excited charmonium decays by flux--tube breaking and the
$\psi^{'}$ anomaly at CDF \vspace{0.8cm} }
\author{Philip R. Page\thanks{E-mail : p.page@physics.oxford.ac.uk} \\
{\small \em Theoretical Physics, University of Oxford, 1 Keble Road,
Oxford OX1
3NP, UK}  \\  \\}
\date{February 1995 \vspace{1.5cm}}
\begin{center}
\maketitle
\begin{abstract}

The hadronic decay of radially and orbitally excited charmonium
above charm threshold by $^3 P_{0}$ pair creation and chromoelectric
flux--tube breaking is discussed in an harmonic oscillator
approximation. We find independent evidence from a study of widths
for a 2S admixture in the predominantly 1D state $\psi(3770)$,
and explore the possibility of
metastable radially excited $2 \; ^{3}P_{0,1,2}$ states
being a source of the anomalously large
production of $\psi^{'}$ at the Tevatron. At least one of them is
expected to be narrow
as a consequence of the existence of nodes in the radial
wave function.

\end{abstract}
\end{center}

\newpage
\section{Introduction}

Recently $\psi^{'}$ enhancement at CDF \cite{mangano94} has generated
considerable
interest.
This may currently be the largest discrepancy between the predictions
of the standard model and experiment, and arises because the
behaviour of QCD in the strongly interacting region is inadequately
understood theoretically.
There are four possible sources of enhanced $\psi^{'}$ discussed in the
literature \cite{cho94,close94i,roy94,cdf} : $2^{-+}$ $c\bar{c}$ below
the $\ds D$ threshold \cite{cho94,close94i}
($D = 0^{-}$ and $\ds = 1^{-}$), radial $\chi$--states (i.e. 2P
charmonia) \cite{close94i,roy94},
hybrid charmonium \cite{close94i} and $\psi^{'}$ with a
$c\bar{c}gg$--component \cite{cdf}.
Investigation indicated that $2^{-+}$ is an unlikely candidate
\cite{cho94,close94i} unless
 $\psi^{'}$ contains significant $^{3}D_{1}$
in its wave function.
Detailed study \cite{roy94} of fusion and fragmentation contributions
to the production of radial $\chi$--states
indicated that they could explain the $\psi^{'}$ excess
if they are very narrow \cite{close94i,roy94} ($\sim$ 1 MeV).
We investigate whether this is likely
or even possible in
a QCD inspired model known to be successful for light hadrons
\cite{kokoski87}, namely the non--relativistic flux--tube
model of Isgur and Paton \cite{paton85},
which is related closely to the phenomenologically successful
\cite{geiger} $^{3} P_{0}$--model \cite{leyaouanc73}.
The extension of earlier work on light mesons \cite{kokoski87}
to charmed hadrons is novel in its own right, and will
form the basis for a subsequent study \cite{close95} of hybrid charmonium.

The strategy is to study first the $D\bar{D}$ width of $\psi(3770)$. This
is the cleanest example \cite{cornell}, being a natural 1D-2S candidate.
The only real uncertainty is the 1D/2S--mixing,
though $e^+e^-$-- and $\gamma$--transitions suggest a dominantly
1D state. We find that the width fits with parameters determined
elsewhere \cite{kokoski87,merlin86,kokoski84,private1,bradley};
and discover that
the next potential $c\bar{c}$ state of higher mass, the 3S \cite{isgur85},
would not enable sensible fitting of the $\psi(3770)$ widths.

It is encouraging that the width of $\psi(3770)$ fits well in this picture.
In fact, it fits in so
well that we find the need to constrain the model parameters by
also fitting energetically
higher lying states, such as $\psi(4040)$, $\psi(4160)$ and $\psi(4415)$,
to experiment.

When the model is applied to calculations of the widths of radial
$\chi$--states the results depend rather critically on the mass of
these states, and
hence the phase space available for their $DD$ and $\ds D$ decays.
The $2^{++}$ decays in D-wave into $DD,\ds D$, while $1^{++}
\rightarrow D \ds$ are likely to be near to threshold.
As anticipated in ref.
\cite{close94i} we find the $2^{++}$ generally quite narrow ($\sim 0.5-5$
MeV) in the range of predicted masses. The $1^{++}$ and $0^{++}$
widths are very sensitive
to the nodes of the radial $\chi$ wave function.
It is possible that they are very narrow ($\sim 1$ MeV), but this
would
be a coincidental
conspiracy, which nevertheless can be attained near the nodes.
We find that for sensible
parameter solutions to $\psi(3770)$, it is likely that some of the
${(0,1,2)}^{++}$ widths are considerably reduced, which is sufficient
for the enhanced production of $\psi^{'}$ at CDF.

The basic structure of the paper is as follows.

In \S 2 we discuss the charmonium system and introduce the flux-tube
model and the parameter values used.
In \S 3 we fit the $\psi(3770)$ to experiment and compare the range of
possible mixing angles
with those obtained from $e^+e^-$ annihilation. We are encouraged
by the consistency of these results, which prompts us to look at the
higher
radially
excited states in \S 4, and observe to which degree their experimental
signatures can be accounted for.
The parameters obtained are used in \S 5
to look at radial $\chi$--states and to examine whether they may be
a source of the
anomalously large production of $\psi^{'}$ at CDF.
We conclude by outlining the experimental consequences.

\section{Outline of $c\bar{c}$ phenomenology}

Earlier work \cite{cornell,bradley,leyaouanc77,kogerler}
on the charmonium system
\cite{novikov78,rosner87,barnes93} is extended by taking into account
flux--tube breaking \cite{kokoski87,geiger,alcock}. We perform
an {\it analytical calculation} of radially and orbitally
excited charmonium decay amplitudes
to $D D$, $\ds D$, $\ds \ds$, $\D \D$, $\Ds \D$, $\Ds \Ds$
and $\dss D$ ($\dss = {(0,1,2)}^{+}$) in a S.H.O. wave function
approximation. This approximation allows comparison with previous work
\cite{leyaouanc73,bradley,leyaouanc77,kogerler} in the $^3 P_{0}$--model,
improves the ability to handle future parameter changes, and is
known to be an excellent for charmonia
and charmed mesons\footnote{
When expanding full numerical wave functions
the overlap with the corresponding radially excited S.H.O. wave function
is typically found to be dominant, e.g. 99.8\% ($D,\ds$),
97.7\% ($\psi(4040)$) and 97.0\% ($\psi(4415)$) \protect\cite{bradley}
for appropriately chosen $\beta$. The nodal positions in wave functions
and decay amplitudes are also well approximated \protect\cite{bradley}.}.
The calculation is performed in the approximation where the inverse radii
$\beta$ of the outgoing mesons are identical. The main reason for this
simplification is that
the $\dss, \D, \Ds, \ds$ and $D$ are expected \cite{kokoski87} to have
similar $\beta$'s, and
that the flux--tube and $^{3}P_{0}$--models correspond closely
in this case (see the next paragraph). Results do not depend
critically
on this
simplification, making it unnecessary to concern ourselves with small
corrections.

The flux--tube model decay amplitude \cite{kokoski87} by pair
creation differs from the familiar $^{3}P_{0}$--model amplitude,
although they coincide in the case when $\beta$ is a
{\it universal constant} for all outgoing mesons.
This is shown in Appendix A, \S \ref{3p0mesonsection}.
In this work, we allow variation of $\beta$. Hence the two models are
not identical, though
their qualitative features are similar \cite{geiger},
allowing comparison with
earlier studies \cite{leyaouanc73,bradley,leyaouanc77,kogerler}.

The only free parameter in the model is the
overall normalization of decays
$\frac{a \tilde{c}}{9 \sqrt{3}} \frac{1}{2} A^{0}_{00} \sqrt{\frac{fb}{\pi}}$
(eqn. \ref{3p0amplitude}, Appendix B). This dimensionless factor is
common with light meson decays
and was phenomenologically found to equal $0.64$ \cite{kokoski87} for
creating light quark pairs with $u,d$ and $s$ flavours\footnote{
\label{3p0strange}Suppressions of 1 \protect\cite{leyaouanc77,kogerler},
$m_u/m_s$ \protect\cite{leyaouanc77} and ${(m_u/m_s)}^2$
\protect\cite{cornell} in amplitude
have been employed for $s\bar{s}$ creation
relative to $u\bar{u},d\bar{d}$ creation. We adopt the first.}. We
adopt it here; thus implying that $c\bar{c}$ widths are
predicted {\it independently}.

{}From the ISGW non-relativistic fit \cite{isgur89} to
spin-averaged meson masses we take
the string tension $b=0.18 \: {\mbox{\rm GeV}}^{2}$, and the
constituent-quark masses
$m_{u,d}=0.33$ GeV, $m_{s}=0.55$ GeV and $m_{c}=1.82$ GeV.
Meson masses are taken to be the PDG masses \cite{pdg94}, and where
not available (which is also the case for $^{3}P_{1} / \, ^{1}P_{1}$
mixing angles) motivated by spectroscopy predictions \cite{isgur85},
adjusted in absolute value relative to known masses.
The procedure for the calculation of widths is outlined in refs.
\cite{kokoski87} and \cite[\S 2]{close94}\footnote{We employ for the
masses $\tilde{M}$ \cite{kokoski87} the values 1.97 GeV ($D,\ds$),
2.07 GeV ($\D,\Ds$),
 2.44 GeV
($\dss$), 3.76 GeV ($\psi(3770)$), 4.015 GeV ($\psi(4040)$), 4.15 GeV
($\psi(4160)$), 4.39 GeV ($\psi(4415)$)}.

The experimental total decay widths in MeV are \cite{pdg94} :
\beqn
\label{3p0experiment}
\barr{llcr}
\psi (3770)  \rightarrow  D D \; \ddag & 23.6 & \pm & 2.7\\
\psi (4040)  \rightarrow  D D \; \ddag, \ds D \; \ddag, \ds \ds \;
\ddag, \D \D & 52.0 & \pm & 10.0 \\
\psi (4160)  \rightarrow  D D, \ds D, \ds \ds, \D \D, \Ds \D&
78.0 & \pm & 20.0 \\
\psi (4415)  \rightarrow  D D, \ds D, \ds \ds, \D \D, \Ds \D,
\Ds \Ds, \dss D & 43 & \pm & 15
\earr
\eeqn
where we indicated the dominant\footnote{
OZI forbidden decays are are expected to be small (e.g. experimental
indicattions are that $\psi(3770) \rightarrow \psi \pi \pi$
has a width of $\sim 20-80$ keV
\protect\cite{yan}). Also, $\psi(4415)$ decays to e.g. $D(2S)D$, $\dss \ds$,
$\Ds(2S)\Ds$ and $\dss_{s} \Ds$ are expected
\protect\cite{isgur85,pdg94} to be
kinematically forbidden.} decay modes considered in this work, and
$\ddag$ indicates modes that have been observed \cite{pdg94}. The
modes
indicated in
eqn. \ref{3p0experiment} are
assumed to dominate over decays to $c\bar{c}\; + \; mesons$.
In the next section we shall calculate these contributions to the
hadronic width in our model, and
sum them up in order to fit them to the total
experimental widths in eqn. \ref{3p0experiment}.

\btablec
\caption{The outgoing meson on--shell CM momentum $p_{B}$ in GeV.}
\label{3p0phasespacetable}
\begin{tabular}{|l|r||l|r|}
\hline 
& $p_{B}$ & & $p_{B}$ \\
\hline \hline 
$\psi (3770) \rightarrow DD       $& 0.26 &
$\psi (4415) \rightarrow DD       $& 1.18 \\ \cline{1-2}

$\psi (4040) \rightarrow DD       $& 0.77 &
$\psi (4415) \rightarrow \ds D    $& 1.06 \\

$\psi (4040) \rightarrow \ds D    $& 0.57 &
$\psi (4415) \rightarrow \ds \ds  $& 0.92 \\

$\psi (4040) \rightarrow \ds \ds  $& 0.21 &
$\psi (4415) \rightarrow \D  \D   $& 1.00 \\

$\psi (4040) \rightarrow \D \D    $& 0.45 &
$\psi (4415) \rightarrow \Ds \D   $& 0.84 \\ \cline{1-2}

$\psi (4160) \rightarrow DD       $& 0.92 &
$\psi (4415) \rightarrow \Ds \Ds  $& 0.65 \\

$\psi (4160) \rightarrow \ds D    $& 0.75 &
$\psi (4415) \rightarrow \dss_{2^{++}} D $&0.44 \\

$\psi (4160) \rightarrow \ds \ds  $& 0.54 &
$\psi (4415) \rightarrow \dss_{1^{+L}} D $&0.52 \\

$\psi (4160) \rightarrow \D  \D   $& 0.67 &
$\psi (4415) \rightarrow \dss_{1^{+H}} D $&0.46 \\

$\psi (4160) \rightarrow \Ds \D   $& 0.41 &
                                   &            \\
\hline 
\end{tabular}
\etablec

\section{Fitting the overall widths to experiment \label{3p0fitsection} }

\subsection{$\psi(3770)$ \label{3p0fittingsection}}

For some time there has been a need to obtain an overall picture of
excited charmonium with modern parameters.
The process $\psi(3770) \rightarrow DD$
produces a $D$--meson of momentum $p_{B} = 0.26$ GeV
(see table \ref{3p0phasespacetable}).
The decay happens far from where the widths vanish,
i.e. from the nodes in the 1D/2S--amplitude, for all phenomenologically
relevant $(\beta_{A},\beta )$. This, the fact there is only
one hadronic decay mode, and that the width has been estimated
successfully before \cite{cornell,ono81},
make it the cleanest charmonium candidate
above charm threshold. We shall now proceed to discuss
this state in more detail.

The $J / \psi$ has unambigously been identified
as dominantly 1S, so the $c\bar{c}$ state $\psi(3770)$ must be a higher
radial or orbital excitation: most probably, on mass alone, a 1D--2S
mixture \cite{isgur85}.
This can independently and non-trivially be  established by
considering its width.
We find that the $\psi(3770)$ width is consistent with it being 1D or 2S,
i.e. with a {\it 1D--2S mixture}. The alternative ``gedanken'' assumption
that $\psi(3770)$ is a 3S or 4S state leads to a contradiction:
this would imply a width to $DD$ too
small to be consistent with experiment (for $\beta_{A},\beta \sim
0.1-1$ GeV). Hence we believe that the 1D-2S nature of $\psi(3770)$
established from width considerations,
being consistent with mass spectroscopy, is significant.

Noting that the $\psi^{'}(3685)$ is predominantly 2S, we expect $\psi(3770)$
to be mainly 1D, with a 2S admixture\footnote{
We neglect 1S admixture for the following reason : Although our
calculation shows that taking $\psi(3770)$ as pure 1S is sensible for
$\beta_{A} \sim 0.1 - 0.5$ GeV and $\beta \sim 0.2 - 0.8$ GeV,
and it is known that 1S mixing in
$\psi(3770)$ can be significant \protect\cite{richard} for spin--dependent
forces; it is small in coupled channel treatments
\protect\cite{cornell,tornqvist,nijmegen}. The latter models
successfully reproduce the experimental 2S admixture.
} as $|\psi(3770) \: \rangle =
\cos \theta \; |1D \rangle + \sin \theta \; |2S \rangle $.
In figure 1 we indicate regions in parameter space where $\psi(3770)$
can be fitted to experiment as a 1D--2S mixture for various $\theta$.
The fact that the
width can be fitted at all is non-trivial, as will be briefly
outlined. If we mix the 1D and 2S decay amplitudes
as a new amplitude $f = \cos \theta \; 1D + \sin \theta \; 2S$,
then $f^{2} \leq {{(1D)}^{2}+{(2S)}^{2}}$; with the maximum of
$|f|$ occuring
at $\tan \theta = 2S/1D$ and the
minimum $f = 0$ at $\tan \theta = - 1D/2S$. Clearly there
is a large range of widths that the mixed state can have as
$\theta$ is varied. There is still a restriction, though:
in regions where ${{(1D)}^{2}+{(2S)}^{2}} < 24 \pm 3$ MeV,
the $\psi(3770)$ total width $f^{2} \leq {{(1D)}^{2}+{(2S)}^{2}}$
cannot be fitted for any $\theta$.
However, there is a large region where $f$
is large enough to reproduce experiment.

We can restrict the parameter space $(\beta_{A},\beta)$
by incorporating {\it theoretical prejudice} about the $\beta$'s.
$D,\ds$ and $\dss$ have $\beta$ in the region\footnote{
\label{3p0theoryfootnote}
It is estimated that $\beta_{D,\ds}$ = 0.39 GeV \cite{isgur89},
0.40 GeV \cite{bradley}, 0.42 GeV \cite{alcock}, 0.49 - 0.55
GeV \cite{kokoski84} or 0.54 - 0.66 GeV \cite{kokoski87};
and $\beta_{\dss}$ = 0.34 GeV \cite{isgur89}, 0.42 - 0.49 GeV
 \cite{kokoski84} or 0.45 - 0.54 GeV \cite{kokoski87} (on average 0.07 GeV
lower than $\beta_{D,\ds}$). For $c\bar{c}$ the estimates are :
$\beta_{1S}$ = 0.66 GeV \cite{isgur89}, 0.75 GeV \cite{merlin86} or
0.57 GeV \cite{private1};
$\beta_{1P}$ = 0.50 GeV \cite{merlin86};
$\beta_{1D}$ = 0.37 GeV \cite{private1}, 0.45 GeV \cite{merlin86}
or 0.48 GeV \cite{alcock}.
} $0.30 - 0.70$ GeV; and $\psi(2S)$ and $\psi(1D)$ have
$\beta_{A} = 0.30 - 0.55$ GeV.
{}From figure 1 we note that
$\psi(3770)$ can be fitted for any $\theta$ in this region of
parameter space, although {\it a sizable 2S admixture} with $\sin \theta
\sim 30 - 45$ \% appears to be
preferred. This will be reinforced when we fit more massive
states in \S \ref{3p0mixingsection}.

It can be estimated\footnote{
The experimental uncertainties in $e^+e^-$--widths \cite{pdg94}
translates into $\sim 4\%$ uncertainty in $\sin \theta$.}
from the observed $e^+e^-$--widths of $\psi^{'}$ and
$\psi(3770)$ that {\it either} $\sin \theta$ = 17\% \cite{yan}
or 23\% \cite{chao} {\it or} $\sin \theta$ = --49\% \cite{yan} or
$\sin \theta$ = --44\% \cite{chao}, employing our convention for $\theta$.
Clearly our analysis prefers $\sin \theta > 0$, concurring with findings
on the detailed analysis of $\psi \pi \pi$ \cite{yan} and $c\bar{c}$
radiative \cite{chao} decays. This also in agreement with
theoretical calculations in the coupled channel formalism, which find
$\sin \theta$ = 17\% \cite{cornell} or 20\%
\cite{tornqvist}.

Given that the overall normalization of decays was taken from
$\it light$ meson spectroscopy as part of a unified study,
in contrast to earlier $c\bar{c}$ calculations
\cite{bradley,leyaouanc77,kogerler,alcock,femi},
we do not regard the disagreement of our estimate with that from
$e^+e^-$--widths as significant. Our determination of the 2S admixture in
$\psi(3770)$ is hence broadly consistent with its
$e^+ e^-$ width and the phenomenology of the $\psi^{'}$.
We thus consider $(\beta_{A},\beta)$
consistent with a realistic and dominant 1D
component in $\psi(3770)$, but with a sizable 2S admixture of $\sim
30\%$ in amplitude.

In summary we conclude that
this model of decay by flux--tube breaking can successfully
account for the $\psi(3770)$ as a 1D--2S mixture in this study of widths,
consistent with {\it both}
the {\it a priori} theoretical estimates for $(\beta_{A},\beta)$ and
the restrictions on $\theta$ from $e^+ e^-$--widths. These
conclusions based on its width are also
in line with mass spectroscopy. In the
next section we shall find that higher states can also be described
with varying degrees of success.

\subsection{Higher mass charmonium \label{3p0mixingsection}}

As far as states more massive than $\psi(3770)$ are concerned, we are led
to identify them as 3S--2D mixtures, or higher radial
excitations, due to the existence of convincing candidates for
1S, 2S and 1D states.
The $\psi(4160)$ and $\psi(4040)$ (which may be either simple
resonances or a collection of states),
should at least contain the 3S and 2D states, otherwise
the $\psi(4415)$ must be 3S or 2D, leading to a contradiction:
the 1S, 2S, 3S, 1D and 2D wave
functions produce $\psi(4415)$ total widths too large to be
consistent with experiment
(for $\beta_{A},\beta \sim 0.2-0.8$).
Hence by default the natural interpretation of $\psi(4160)$ and
$\psi(4040)$ is that they are
3S--2D mixtures. They can be fitted consistently with experiment in the
phenomenologically relevant region $\beta_{A},\beta \sim 0.3 - 0.6$.
Also, $\psi(4415)$ can be fitted
as 4S for $\beta_{A} \sim 0.3 - 0.6$ GeV and $\beta \sim 0.4 - 0.6$ GeV.
We shall hence adopt\footnote{
We assume no mixing for higher charmonium states for simplicity :
it is generally considered to be small \cite{cornell}, e.g. $\sim 10\%$
\cite{tornqvist} in coupled channel models.} these assignments,
which are in agreement with mass spectroscopy
\cite{isgur85,perantonis90}.

We expect the $\psi(4160)$ to be the easiest to describe. This is
because there is an experimental constraint on the width ratios of
$\psi(4040)$ \cite{leyaouanc77} (see below), casting doubt on
whether it is conventional
$c\bar{c}$. Also, the $\psi(4415)$ has numerous decays modes
(including decays to $\dss D$),
making its narrowness \cite{leyaouanc77} potentially difficult to understand.

It is theoretically expected that the magnitudes of $\beta_{A}$ for
higher excited states are
smaller than those of lower states (e.g. $\psi(3770)$ in this case)
by $\lapprox 0.1$ GeV
\cite{kokoski87}. Allowing for this, we find\footnote{
This result holds for $\psi(4040)$ as 3S, and to a lesser extent
as 2D; for $\psi(4160)$ as 3S, and to a lesser extent as 2D; and for a 4S
$\psi(4415)$.
} that all three of the higher states can
be accommodated with a 2S admixture  in
$\psi (3770)$ of $\sim 30 - 45$ \% in amplitude. It can be seen by comparing
figures 1 and 2.

If we choose parameters consistent with a sizable 2S component
in $\psi(3770)$, figure 2 indicates that it is possible to fit
$\psi(4160)$ as 3S, and to a lesser extent as
2D, in a wide range of theoretically acceptable
$(\beta_{A},\beta)$. If we require $\beta$ near to the ISGW value
\cite{isgur89},
$(\beta_{A},\beta)$ = (0.30,0.39) GeV
gives a good fit for most states.
Some of these values also allow fits of $\psi(4415)$
(see tables \ref{3p0widthtable} and \ref{3p0widthtable1}),
e.g. $(\beta_{A},\beta)$ = \underline{(0.42,0.52)} GeV, where
we find excellent agreement.
The point
$(\beta_{A},\beta)$ = (0.50,0.54) GeV,
which fits $\psi(4040)$ in addition (see below), also lies in this region.

\btablec
\caption{A selection of widths in MeV
{\it fitting experiment} (see \S \protect\ref{3p0fitsection}).
We assume a common $\beta$
(in GeV) in order to reduce parameters. We allow
$\psi(3770)$ to have $\beta_{A}^{\psi(3770)} = \beta_{A} + 0.1$ GeV
(see \S \protect\ref{3p0mixingsection}), and a 1D-2S mixing angle
$\theta$,
in the decay
$\psi(3770) \rightarrow DD$.
We indicate $\psi(4160)$ as either 3S or 2D, and the widths of the
various decay modes $\psi(4160) \rightarrow
DD, \ds D, \ds \ds$ (P- and F-wave), $\D \D, \Ds \D$.}
\label{3p0widthtable}
\begin{tabular}{|r|r||r|r||r|r|r|r|r|r|r|r|}
\hline 
\multicolumn{2}{|c||}{} & \multicolumn{2}{c||}{$\psi(3770)$} &
\multicolumn{3}{c}{$\psi(4160)$} & \multicolumn{2}{c}{$\ds \ds$} &
\multicolumn{3}{c|}{} \\
$\beta_{A}$ & $\beta$ & $\sin \theta$ & Total &  & $DD$ &
$\ds D$ & P & F & $\D \D$ & $\Ds \D$ &
Total \\
\hline \hline 
0.50 & 0.54 & 0.3 & 23 & 2D & 32 &  1 &  6 & 20 &  4 & 11 & 74 \\
0.50 & 0.54 & 0.3 & 23 & 3S & 13 &  4 & 34 &  0 &  1 & 20 & 72 \\
0.42 & 0.52 & 0.4 & 24 & 3S & 17 & 44 &  3 &  0 &  1 &  2 & 67 \\
0.30 & 0.39 & 0.5 & 21 & 3S &  1 & 22 & 52 &  0 &  5 &  1 & 81 \\
\hline 
\end{tabular}

\caption{As in table \protect\ref{3p0widthtable} but for $\psi(4415)$
decaying to $DD$, $\ds D$, $\ds \ds$ (P-wave),
$\D \D$, $\Ds \D$, $\Ds \Ds$ (P-wave), $\dss_{2^{++}}$,
$\dss_{1^{+L}}$ (S- and D-wave), $\dss_{1^{+H}}$ (S- and D-wave).
Here {\it L,H} indicates the low and high $J^{P}=1^{+}$ mass states
respectively, with a $^{3}P_{1} / \, ^{1}P_{1}$--mixing angle of
$\tilde{\theta} = -41^{o}$
\protect\cite{isgur85} (see eqn. \protect\ref{3p0tranform}, Appendix B).
We take the $\dss_{1^{+H}}$ mass to be 2.45 GeV.}
\label{3p0widthtable1}
\begin{tabular}{|r|r||r|r|r||r|r|r||r|r|r|r|r||r|}
\hline 
$\beta_{A}$ & $\beta$ & $DD$ & $\ds D$ & $\ds \ds$ & $\D \D$ &
$\Ds \D$ & $\Ds \Ds$ &
$\dss_{2^{++}}$ &
\multicolumn{2}{c|}{$\dss_{1^{+L}}$} &
\multicolumn{2}{c||}{$\dss_{1^{+H}}$} &
Total \\
& & & & & & & & & S & D & S & D & \\
\hline \hline 
0.50 & 0.54 &  1 &  3 &  1 &  1 &  2 &  0 & 14 &  0 & 17 &  1 &  1 & 41 \\
0.42 & 0.52 &  8 & 22 & 11 &  1 &  0 &  0 &  1 &  0 &  1 &  1 &  0 & 45 \\
\hline 
\end{tabular}
\etablec

We now proceed to discuss the $\psi(4040)$. The issue of whether
it can be understood as ordinary
charmonium has historically been controversial, due to its anomalously
large $\ds \ds$ branching ratio.
The large experimental ratios \cite{pdg94}
$R_{1} \equiv \Gamma (\ds D) \; /$ $\Gamma (D D)$ = 5 - 20
and $R_{2} \equiv$ $\Gamma (\ds \ds) \; /$ $\Gamma (\ds D)$ = 1 - 2.5
can conceivably be
explained by looking for nodes (zero's) \cite{cornell}
in the decay amplitudes corresponding
to decays into $DD$ and $\ds D$.
This can arise if the kinematics of the decay are
controlled by nodes arising from the radial wave functions of the
3S or 2D states.
There is thus the possibility of the $DD$ and $\ds D$ amplitudes being near
enough to two\footnote{
We showed that there exits no $(\beta_{A},\beta)$ for which for which the
momenta $p_{B} = 0.77$ GeV ($D D$) and $p_{B} = 0.57$ GeV ($\ds D$)
lie near different nodes in the decay amplitude.
} different nodes or to the same node \cite{leyaouanc77}.
We proceed to seek cases
$(\beta_{A},\beta )$ where $R_{1}$
and $R_{2}$ are consistent with experiment.
Although $R_{1,2}$ can be fitted for $\psi(4040)$
as 2D with $(\beta_{A},\beta) \sim (0.3,0.2)$, the total width is too
small, leading us to the conclusion that a consistent picture of
$\psi(4040)$ can {\it only} be obtained with it as 3S.
Our search provides three areas in parameter space where
the $\psi(4040)$ may be
realized as 3S (see table \ref{3p04040table} and figure 2).
If in addition we require the $\psi(3770)$ and $\psi(4160)$
to be consistent with experiment, $\psi(4160)$ to
be 2D (as a consequence of $\psi(4040)$ being 3S), and the parameters
to be consistent with theoretical estimates, we are restricted to
$(\beta_{A},\beta )$ around \underline{(0.50,0.54)} GeV.
This region represents the only area where we can fit {\it all}
experimental data on
excited charmonium decays, including the $\psi(4160)$ and $\psi(4415)$, with
the $\psi(3770)$ a 1D-2S mixture as usual
(see tables \protect\ref{3p0widthtable1}
and \protect\ref{3p04040table}). The region also appears to be
related to the one found by Le Yaouanc {\it et al.}
\cite{leyaouanc77}, who also identified\footnote{
They obtained $\beta_{A} = \beta = 0.44$ GeV, but with old parameter
values and a mass formula neglecting the difference between $m_{c}$
and $m_{u}$.} $\psi(4040)$ as 3S.
This implies that it is possible to fit
$\psi(4040)$ consistent with experiment without
contradicting the phenomenology of the other $\psi$--states, although it
{\it severely} constrains the parameters, requiring considerable
coincidence of parameters \cite{kogerler}
and fine tuning of $\beta_{A}$ to $\sim 1 - 3$ \%,
in order to obtain the correct branching ratios.

The fits obtained for $(\beta_{A},\beta)$ = (0.42,0.52) and (0.50,0.54) GeV
above are remarkable in the sense that we can
accurately fit {\it three} amplitudes (i.e. $\psi(3770)$, $\psi(4160)$
and $\psi(4415)$)
to experiment with {\it two} parameters $\beta_{A}$ and $\beta$, with the
{\it additional} conclusion that $\psi(4040)$ can possibly be
fitted, and furthermore that the assigned nS, nD classifications are
consistent
with mass spectroscopy.

In summary we conclude that the higher states may
also be accommodated in this model. We have thus established another
rung on the
$c\bar{c}$ ladder, highlighting the difficulties associated with
$\psi(4040)$ and $\psi(4415)$.

\btablec
\caption{Total widths in MeV at specific
$(\beta_{A},\beta)$ (in GeV)
{\it providing an excellent fit to $\psi(4040)$} (see \S
\protect\ref{3p0mixingsection}).
$\psi(3770)$ is fitted with a 1D-2S
mixing angle $\theta$ assuming
$\beta_{A}^{\psi(3770)} = \beta_{A} + 0.1$ GeV. The $\psi(4040)$ width ratios
$R_{1,2}$ defined
in \protect\S \protect\ref{3p0mixingsection} are indicated along with
the widths of the various decay modes.
$\psi(4040)$ is constained to be 3S, implying that
$\psi(4160)$
should be 2D. All decays are in P-wave.}
\label{3p04040table}
\begin{tabular}{|r|r||r|r|r|r|r|r|r||r|r||r|}
\hline 
\multicolumn{2}{|c||}{} & \multicolumn{7}{c||}{$\psi(4040)$} &
\multicolumn{2}{c||}{$\psi(3770)$} & $\psi(4160)$ \\
$\beta_{A}$ & $\beta$ & $DD$ & $\ds D$ & $\ds \ds$ & $\D \D$ & Total
&$R_{1}$ & $R_{2}$ & $\sin \theta$ & Total &
Total \\
\hline \hline 
0.50 & 0.54 &  2 & 15 & 23 &  5 & 45 & 9 & 1.5 & 0.3  & 23 & 74 \\
0.24 & 0.23 &  4 & 26 & 34 &  2 & 66 & 6 & 1.3 & 0.1  & 22 & 41 \\
0.50 & 0.20 &  1 & 14 & 29 &  8 & 52 &15 & 2.0 &      &    &    \\
\hline 
\end{tabular}
\etablec

\section{$2P$ charmonia and the $\psi^{'}$ anomaly
at CDF}

The experimental production rate of $\psi^{'}$ is 30 times larger than
theoretical estimates \cite{mangano94} at large transverse momentum in
$p\bar{p}$
collisions at the Tevatron.
One proposed source \cite{close94i,roy94} is the production of
2P charmonia followed by their radiative decay
$ 2 \: ^{2S+1} P_{J}  \rightarrow 2 \; ^{3}S_{1} + \gamma$,
enhancing the $\psi^{'}$ rate.
This can only happen if the 2P charmonia have
small hadronic widths. These states are expected
\cite{isgur85,perantonis90}
to lie above
the $DD$--threshold (3.73 GeV), probably above the $\ds D$--threshold
(3.87 GeV), and possibly above the $\D \D$--threshold (3.94 GeV)
but beneath the $\ds \ds$--threshold (4.02 GeV). In these
circumstances we find that their hadronic widths may be small.

If the 2P
states lie {\it below} the $\ds D$--threshold (say at 3.85 GeV), then the
$ 2 \: ^{3}P_{2}$ and $2 \: ^{3,1} P_{1}$ are narrow :
$ 2 \: ^{3}P_{2}$ (because of D-wave phase space) has width 0-4 MeV,
and $ 2 \: ^{3,1}P_{1}$ is narrow because only decays to $\ds D$ are allowed.

If the 2P states lie {\it above} the $\ds D$--threshold the
$2 \: ^{3}P_{2}$ has the unique property that the
total width remains small (because of D-wave phase space) at
0 - 7 MeV (3.90 GeV) and 1 - 14 MeV (3.95 GeV).
Figure 3 indicates that in the theoretically reasonable region
where $(\beta_{A},\beta) \sim (0.3,0.4) - (0.4,0.6)$ GeV
these widths can be $\lapprox 1$ MeV
if $(\beta_{A},\beta)$ is tuned to only $\sim$ (15,20) \%.
This is of special interest for the resolution of the CDF anomaly.
The $ 2 \: ^{3,1}P_{1}$ and $2 \: ^{3}P_{0}$ states
only have small widths if the kinematics of the decay cause the
amplitude to be controlled by the nodes arising from the radial
wave functions of the 2P states
(otherwise the amplitude can be substantial, even near thresholds).

\begin{itemize}

\item For decays of $ 2 \: ^{3,1}P_{1}$ the S-wave nodes occur in
a line of
parameter space $(\beta_{A},\beta) \sim (0.3,0.4) - (0.45,0.6)$ GeV
coinciding with narrow $2 \: ^{3}P_{2}$ (see figure 3).
It is fascinating that this region coincides with D-wave nodes,
reinforcing the narrowness of $ 2 \: ^{3,1}P_{1}$.

\item For decays of $2 \: ^{3}P_{0}$ the S-wave nodal line occurs for
$(\beta_{A},\beta) \sim (0.4,0.3) - (0.5,0.6)$ GeV
(see figure 3).
Estimates \cite{isgur85}
suggest this to be the lightest 2P state, and hence possibly narrow
(i.e. below the $\D \D$--threshold, which ensures that
$2 \: ^{3}P_{0} \rightarrow \D \D$ does not overshadow the decay).

\end{itemize}

Hence we expect {\it at least one} of $2 \: ^{3}P_{2}$,
$2 \: ^{3}P_{1}$ {\it or} $2 \: ^{3}P_{0}$
to be narrow; the $2 \: ^{3}P_{2}$ being most likely, since
obtaining $2 \: ^{3}P_{0}$ and $2 \: ^{3}P_{1}$ widths of
$\leq 1$ MeV requires fine tuning $(\beta_{A},\beta)$ respectively
to $\sim$ (4,7) \% and $\sim$ (1,1) \% in figure 3.
Theoretically we expext $\beta_{2P} \sim 0.3 - 0.45$ GeV, tantilizingly
in accord with the values where we expect nodes to occur.

Both the best regions in \S \ref{3p0fitsection} fitting experiment
(i.e. near $(\beta_{A},\beta) = (0.42,0.52)$
and $(0.50,0.54)$ GeV) can be consistent with one of the 2P states being
very narrow ($\sim 1$ MeV), depending sensitively on their masses.

If an amplitude lies near a node, it becomes especially
sensitive to differential amounts of phase space
e.g. for $D^{0}{\bar{D}}^{0}$ and $D^{+}{\bar{D}}^{-}$ final states.
This sensitivity may be decreased if other small decay modes far from a
node swamp the decay.
Since $\delta p_{B} \sim 1/p_{B}$,
the change in available phase space is largest when the
decay is nearest to threshold, so that
amplitude variations\footnote{We have not considered these
effects in \S \protect\ref{3p0fittingsection}.
They are expected to be largest for $\psi(3770) \rightarrow DD$, because
it is near threshold (see table \protect\ref{3p0phasespacetable}), and
the dominant
hadronic decay mode. However for realistic $(\beta_{A},\beta)$ it is far
from a node, with typically 5 - 15 \% variation in width due to varying
the phase space of decays to $D^{0}{\bar{D}}^{0}$ and $D^{+}{\bar{D}}^{-}$.
So the effect is small compared to variations in
$(\beta_{A},\beta)$.} are pronounced near thresholds.
Hence, when 2P masses are known experimentally, these effects should
be
incorporated
near the various thresholds if there is nodal suppression of an amplitude.

\section{Conclusions}

Our results from hadronic decays of charmonium are as follows :

\begin{itemize}

\item $\psi (3770) \rightarrow D D$ can be understood with the
$\psi (3770)$ as a mixture of $2 \: ^{3} S_{1}$ and $1 \: ^{3} D_{1}$.
We find that the width fits with parameters determined
elsewhere \cite{kokoski87,merlin86,kokoski84,private1}; and discover that
the next state of higher mass, the 3S \cite{isgur85},
would not enable sensible fitting.
We favour a sizable 2S admixture in
$\psi(3770)$ ($\sim 30\%$ in amplitude),
consistent with $e^+e^-$--annihilation.
This is also in observational agreement with higher states.

\item We find further encouragement
that the widths of $\psi(4160)$ and $\psi(4040)$ fit well
if they are 3S--2D mixtures. We note that they do not admit solutions
as 2S-1D states.

\item $\psi (4160) \rightarrow D D, \ds D, \ds \ds, \D \D,
\Ds \D$ can be
understood with the $\psi (4160)$ as either $3 ^{3} S_{1}$ or $2 ^{3} D_{1}$,
depending on the preferred value of $(\beta_{A},\beta)$ in
\S \ref{3p0fitsection}.
Study of the branching ratios of this state at a $\tau$-charm factory
could be a rather critical test of its wave function composition.

\item $\psi (4040)$ cannot be understood as $2 \: ^{3} D_{1}$,
due to an inability to reproduce simultaneously the total width and
correct branching ratios into
$D D, \ds D, \ds \ds$. These are experimentally (neglecting statistical
uncertainties) 1 : 8 : 14. Remarkably,
the state can be understood as $3 \: ^{3} S_{1}$ in three regions of
parameter space (see table \ref{3p04040table}).
This, however, severely constrains parameters.
Previous interpretations of $\psi(4040)$ as $c\bar{c}$
\cite{bradley} may hence now be invalid due
to a change in preferred parameters.
This leaves open the possibility of
$\psi (4040)$ being a $\ds \ds$ molecule
\cite{novikov78} or a threshold effect \cite{barbieri75}
(the state is 10 - 30 MeV above the $\ds \ds$ threshold with a
width of 40 - 60 MeV, and can hence not be treated well in the narrow
resonance approximation). There may be several $1^{--}$ states in the
4.0 - 4.3 GeV region, due to the possible additional existence
of a hybrid meson of mass $\sim 4.2$ GeV \cite{perantonis90,swanson94},
but whose $e^+e^-$ production is suppressed due to the radial wave
function
vanishing at the origin \cite{perantonis90}.

\item The $\psi(4415)$ is pinned down to having a
4S wave function. Its total width tends to be small enough
only in restricted regions, suggesting that the narrowness
\cite{leyaouanc77} of $\psi(4415)$ remains an interesting issue\footnote{
The narrowness can be increased by increasing the radial excitation.
Studies of the $\psi(4415)$ width
\cite{leyaouanc77,kogerler,alcock} and mass
\cite{tornqvist} tend to find a 4S assignment.
A 5S assignment (with an accompanying 4S $\psi(4160)$) has, however,
been suggested \cite{chao,femi}. This possibility is found
to be consistent with experimental widths by our study, and can as
such not be ruled out.}.
It is not easy \cite{femi} to identify dominant {\it decay modes}
without fixing $(\beta_{A},\beta)$ and the degree of mixing,
as was effectively done in the literature
\cite{leyaouanc77,kogerler,bradley}, in contrast
to earlier expectations \cite{leyaouanc77,kogerler} where $\dss_{2^{++}} D$
was indicated as possibly significant.
\end{itemize}

Where energetically allowed, decays to $\D,\Ds$ are suppressed by a
flavour factor of two, relative to decays to $D,\ds$.
This, together with P-wave phase space, conspires to make the
branching ratios
of $\psi(4040)$, $\psi(4160)$ and $\psi(4415)$ to $D$ and $\ds$
consistently larger than those to $\D$ and $\Ds$. This is violated
for $\psi(4040) \rightarrow \D \D$,
where the lack of a corresponding experimental width estimate may indicate
additional suppression of $s\bar{s}$ creation \cite{cornell,leyaouanc77}
(see table \ref{3p04040table} and footnote \ref{3p0strange}).

Radial $\chi$--states may be
the source of a significant $\psi^{'}$ cross--section at the Tevatron
if their widths are narrow.
Their widths depend rather critically on their masses, and
hence the phase space available for their $DD$ and $\ds D$ decays.
The $2^{++}$ decays in D-wave into $DD,\ds D$, while $1^{++}$ decays to
$D \ds$ are likely to be near to threshold. As anticipated in ref.
\cite{close94i} we find the $2^{++}$ generally quite narrow ($\sim 0.5-5$
MeV) in the range of predicted masses, with the narrower widths
corresponding to D-wave nodal suppression.
The $1^{++}$ and $0^{++}$ widths are very sensitive
to the nodes of the radial $\chi$ wave function.
It is possible that they are very narrow ($\sim 1$ MeV), but this
would
be a coincidental
conspiracy. For $0^{++}$ below the $\D \D$--threshold, nodal suppression
is quite consistent with $\psi(3770)$ decays. We thus find that for sensible
parameter solutions to $\psi(3770)$, it is likely that some of the
${(0,1,2)}^{++}$ widths are considerably reduced.
This would suggest placing emphasis on searching for the
$2\: ^{3}P_{2}$ in events containing $\psi^{'}$. We predict that if
its hadronic width is found to
be $\sim$ 1 MeV, then the $2 \: ^{3,1}P_{1}$ may also be found to
narrow (depending sensitively on their mass).


The overall consistency of excited charmonium with observation
vindicates the expectation that the pair creation amplitude is
similar for charmonium and light meson decays (in each case light
quarks are created). In view of the slightly exaggerated 2S admixture
we obtain in $\psi(3770)$, this amplitude may have to be adjusted
slightly.
This procedure would however not affect the nodes in the $\chi$
decay amplitude, leaving the $(\beta_{A},\beta)$ where $\chi$--states are
narrow unaltered.

We highlight the need for improved data on
excited charmonia : to confirm whether they are true resonances,
to study their branching ratios into various channels and to discover
whether some may be narrow. Rather clear signals, in particular in
$e^{+}e^{-}$--annihilation, may also reveal hybrids and molecules.

\vspace{0.7cm}
{\noindent \Large \bf Acknowledgements\\ \vspace{-0.3cm} }

Thanks to J.M. Richard, K. Lane, S.F. Tuan and A. Wambach for helpful
criticism. Special thanks goes to F.E. Close
for discussions, motivation and comments. This work
was supported in part by the Universities of Cape
Town and Oxford.

\setcounter{section}{0}
\renewcommand{\thesection}{\Alph{section}}
\section{{\sc \bf Appendix} : The coincidence of flux-tube and
$^{3}P_{0}$--model decay amplitudes
in limiting cases}

For meson decays to mesons
the $^{3}P_{0}$ and the flux-tube models
coincide in the limit of an infinitely thick flux--tube, i.e.
where the pair creation amplitude is constant all over space. This is
indicated explicitly in \S \ref{3p0a1}. It is
possible to make a stronger statement : {\it The $^{3}P_{0}$ and
flux-tube models coincide when $\beta$ is a {\rm universal constant}
for all outgoing mesons, even if the flux--tube has finite
thickness}.
This is demonstrated in \S
\ref{3p0mesonsection}.

We focus on two initial quarks of mass $M$, with pair creation of
quarks of mass $m$. The decay amplitude of an initial meson A
into final mesons
B and C can be shown \cite[Appendix A]{close94}
in the rest frame of A (where $\bp_{A}=0$) to be given by
\beqna
\label{3p0master}
\lefteqn{ - \frac{a \tilde{c}}{9 \sqrt{3}} (2 \pi  )^{3}
\delta^{3} (\bp_{B} + \bp_{C}) \frac{i}{2} Tr(A^{T}BC)_{flavour}
Tr(A^{T}B \bsigma^{T} C)_{spin} \cdot} \nonumber  \\
& &  \eqntimes \int d^{3} \br_{A} \, d^{3} \by \,
\psi_{A}(\br_{A}) \exp (i \frac{M}{m+M} \bp_{B}
\cdot \br_{A}) \gamma (\br_{A},\by_{\perp})  \nonumber \\
& &  \eqntimes (i \bnabla_{\br_{B}} +
i \bnabla_{\br_{C}} + \frac{2 m}{m+M} \bp_{B})
\psi_{B}^{\ast}(\br_{B}) \psi_{C}^{\ast}(\br_{C})
\; + \; (B \leftrightarrow C)
\eeqna

Here $(B \leftrightarrow C)$ indicates a term obtained by interchanging
the flavour and spin matrices
$B \leftrightarrow C$
and momenta $\bp_{B} \leftrightarrow \bp_{C}$ in the first term in
eqn. \ref{3p0master}.
For $c \bar{c}$ decaying by the creation of a $u\bar{u}$, $d\bar{d}$ or
$s\bar{s}$ pair,
only one of the terms contribute (due to $Tr(A^{T}BC)_{flavour}$); and the
space part of the term differs by a sign from the space part of the
displayed term in eqn. \ref{3p0master}. So from now on it is
sufficient just to consider the the displayed term.

The pair creation position $\by$ \cite{kokoski87} is measured relative
to the CM of the initial quarks, and
 $\by_{\perp} \equiv -(\by \times \hat{\br}_{A})
\times \hat{\br}_{A}$ is a perpendicular ``component'' of $\by$ to
the initial $Q\bar{Q}$--axis $\br_{A}$.
The Q\={q} - axes of the final states B and C
are $\br_{B} = \br_{A}/2 + \by$ and $\br_{C} = \br_{A}/2 - \by$
respectively \cite{kokoski87}.

\subsection{Spatially constant pair creation \label{3p0a1}}

The pair creation amplitude (or {\it flux-tube overlap}) is

\beqn
\label{3p0mesondec}
\gamma (\br_{A},\by_{\perp}) = A^{0}_{00} \sqrt{\frac{fb}{\pi}}
 \exp (-\frac{fb}{2} \by^{2}_{\perp})
\eeqn
The thickness of the flux-tube is related inversely to $f$.
A detailed discussion of these quantities and the structure of
eqn. \ref{3p0mesondec} may be found in ref. \cite[eqn. A21]{kokoski87}
and ref. \cite{perantonis87}.
The estimated values $f=1.1$ and $A^{0}_{00}=1.0$
\cite{close94,perantonis87} are used in this work.
The infinitely thick flux-tube with $f=0$ corresponds to the
$^{3}P_{0}$--model \cite{leyaouanc77}.
In this case $\gamma (\br_{A},\by_{\perp})$ is
a constant (which should be normalized to be non-zero), and
the $q\bar{q}$--pair is created with uniform amplitude
anywhere in space. Fourier transforming (a relevant part of) eqn.
\ref{3p0master} yields

\beqna
\label{3p0fourier}
\lefteqn{\eqnneghspace \frac{-i}{2} (2 \pi)^{3} \delta^{3}
(\bp_{B} + \bp_{C})
\; \be_{\sigma} \cdot \int d^{3} \br_{A} \, d^{3} \by \,
\psi_{A}(\br_{A}) \exp (i \frac{M}{m+M} \bp_{B}
\cdot \br_{A}) \; } \nonumber \\
& & \eqntimes (i \bnabla_{\br_{B}} + i \bnabla_{\br_{C}} + \frac{2
m}{m+M} \bp_{B}) \;
\psi_{B}^{\ast}(\br_{B}) \psi_{C}^{\ast}(\br_{C}) =  \nonumber
\eeqna
\intereqnvspace \beqn
-i \delta^{3} (\bp_{B} + \bp_{C}) \; \be_{\sigma} \cdot
\int d^{3} \bk_{A} \:
(\bk_{A} + \bp_{B}) \: \psi_{A}(\bk_{A}) \,
\psi_{B}^{\ast}(\bk_{A} + \frac{M}{m+M} \bp_{B}) \,
\psi_{C}^{\ast}(\bk_{A} + \frac{M}{m+M} \bp_{B})
\eeqn
where we defined $\psi_{A,B,C}(\br) = {(2 \pi)}^{-3} \int
d^{3} \bk \: \exp (i \bk \cdot \br) \psi_{A,B,C}(\bk)$.

Taking the $^{3}P_{0}$--model amplitude \cite[p. 120]{close79},
with the $q\bar{q}$--pair created with equal but opposite momenta

\[
\frac{-i}{2} \sqrt{\frac{4 \pi}{3}} \int d^{3} \bk_{1} \:
d^{3} \bk_{2} \: d^{3} \bk_{3} \: d^{3} \bk_{4} \:
\delta^{3} (\bk_{1} + \bk_{2} - \bp_{A}) \; \delta^{3}
(\bk_{1} + \bk_{3} - \bp_{B}) \;
\delta^{3} (\bk_{2} + \bk_{4} - \bp_{C}) \; \delta^{3}
(\bk_{3} + \bk_{4}) \;
\]
\intereqnvspace
\beqn
\eqntimes Y_{1 \sigma} (\bk_{3} - \bk_{4})
\, \psi_{A}(\frac{\bk_{2}-\bk_{1}}{2}) \,
\psi_{B}^{\ast}(\frac{M \bk_{3} - m \bk_{1}}{m+M}) \,
\psi_{C}^{\ast}(\frac{m \bk_{2} - M \bk_{4}}{m+M}) \,
\eeqn
we can see that it equals the last line of eqn. \ref{3p0fourier}
\cite{bradley} (with
$\bp_{A} = 0$) when we make a change of variables to $\bk_{A} =
(\bk_{1} - \bk_{2})/2$
and $\bk_{CM} = \bk_{1} + \bk_{2}$. Here
$k Y_{1 \sigma} (\bk ) = \sqrt{\frac{3}{4 \pi}} \: \be_{\sigma} \cdot
\bk$ was used.
We have thus connected the flux-tube and $^{3} P_{0}$--model
notation explicitly; showing that {\it the two models coincide for an
infinitely thick flux--tube}; and
verifying the expectation that {\it a zero momentum $q\bar{q}$--pair
corresponds to complete freedom in creating the pair anywhere in space}.

\subsection{Outgoing mesons of equal ``size'' \label{3p0mesonsection}}

The incoming and outgoing mesons have S.H.O.\ wave functions with
inverse radii $\beta_{A}$ and $\beta$ respectively. We assume the outgoing
$\{ D,\ds,\D,\Ds \}$ and $\dss$ wave functions to have angular momentum L=0
and L=1 respectively. In the lowest radial state the wave functions
are:

\beqn
\psi_{L}(\br) = {\cal N}_{L} r^{L} Y_{L M_{L}} (\hat{\br})
\exp (- {\beta_{L}^{2}r^{2}}/2)
\eqnhspace {\cal N}_{L} = \frac{2 \beta_{L}^{3/2}}{\pi^{1/4}}
\{ 1,\sqrt{\frac{2}{3}} \beta_{L}, \frac{2}{\sqrt{15}} \beta_{L}^{2} \}
\label{3p0ground}
\eeqn
All wave functions are properly normalized (with the brackets referring to
S, P and D states respectively). Performing the
differentiation in eqn. \ref{3p0master} gives for $S+S$ final states
\beqna
\label{3p0finalss}
\lefteqn{ \be^{\ast}_{\sigma} \cdot (i \bnabla_{\br_{B}} + i
\bnabla_{\br_{C}} + \frac{2 m}{m+M}
\bp_{B}) \: \psi_{S}^{\ast}(\br_{B}) \psi_{S}^{\ast}(\br_{C})
= } \nonumber \\  & & {\cal N}_{S}^{2} \exp (-{\beta}^{2}
(\frac{\br_{A}^{2}}{4}+\by^{2})) \;
(- i {\beta}^{2} \br_{A} + \frac{2 m}{m+M} \bp_{B} )
\eeqna
and now proceeding to perform the y-integration gives
\beqna
\int d^{3} \by \gamma (\br_{A},\by_{\perp}) \times (eqn. \, \ref{3p0finalss})
& = &  A^{0}_{00} \sqrt{\frac{fb}{\pi}} {\cal N}_{S}^{2} \: \bar{\gamma}^{0}
\; (- i {\beta}^{2} \br_{A} + \frac{2 m}{m+M} \bp_{B} ) \;
\exp (- \frac{{\beta}^{2} \br_{A}^{2}}{4}) \\ \nonumber
\bar{\gamma}^{0} & \equiv &
\int d^{3} \by \; \exp (-\beta^{2} \by^{2} - \frac{fb}{2} \by^{2}_{\perp})
= \frac{\pi^{3/2}}{\beta (\beta^{2} + fb/2)}
\eeqna

Similarly for $P+S$ final states
\beqna
\label{3p0finalps}
\lefteqn{(i \bnabla_{\br_{B}} + i \bnabla_{\br_{C}} + \frac{2 m}{m+M}
\bp_{B}) \:
\psi_{P}^{\ast}(\br_{B}) \psi_{S}^{\ast}(\br_{C})
= {\cal N}_{S} {\cal N}_{P} \exp (-{\beta}^{2}
(\frac{\br_{A}^{2}}{4}+\by^{2}) )} \nonumber \\
& & \eqntimes
[i \sqrt{\frac{3}{4 \pi}} \be^{\ast}_{M_{L}^{B}} +
(\frac{r_{A}}{2} Y_{1 M_{L}}^{\ast} (\hat{\br}_{A})
+ y Y_{1 M_{L}}^{\ast} (\hat{\by}) ) \:
(- i {\beta}^{2} \br_{A} + \frac{2 m}{m+M} \bp_{B} )]
\eeqna
and performing the y-integration, terms linear in $\by$
(in eqn. \ref{3p0finalps}) vanish
\[
\int d^{3} \by \gamma (\br_{A},\by_{\perp}) \times (eqn. \, \ref{3p0finalps})
 = \] \intereqnvspace
\beqn
\label{3p0integration}
A^{0}_{00} \sqrt{\frac{fb}{\pi}} {\cal N}_{P} {\cal N}_{S} \:
\bar{\gamma}^{0} \;
[i \sqrt{\frac{3}{4 \pi}} \be^{\ast}_{M_{L}^{B}} +
\frac{r_{A}}{2} Y_{1 M_{L}}^{\ast} (\hat{\br}_{A}) \;
(- i {\beta}^{2} \br_{A} + \frac{2 m}{m+M} \bp_{B} )]
\exp (- \frac{\beta^{2} \br_{A}^{2}}{4})
\eeqn

Since the $^{3} P_{0}$--model \cite{leyaouanc73} corresponds to taking $f=0$
in the amplitude above, we see that the flux-tube model amplitude
differs from the $^{3} P_{0}$--model amplitude by a factor of
$1+fb/2 \beta^{2}$
for both {\it S + S} and {\it P + S} final states.
This explains the systematics of an earlier calculation
\cite[table II : compare columns 2 and 3]{kokoski87} by Isgur and Kokoski,
with $\beta=0.4$ throughout. Hence
in the approximation where (the final state) $\beta$ is constant
throughout for
all mesons, {\it the flux-tube and $^{3} P_{0}$--models yield
identical amplitudes} when normalized to experiment.

The reason why Isgur and Kokoski's results for the $^{3} P_{0}$ and
flux-tube models are slightly different, is because they use
a cigar-shaped overlap, which modifies the infinite cylinder overlap of
eqn. \ref{3p0mesondec} longitudinally away from the $Q\bar{Q}$--axis.
The small deviation of their results for the two models from
one another ($\sim 5 \%$ in width) demonstrates that this overlap,
although physically
reasonable, makes little quantitative difference on the scale of
model errors, and is hence
safely neglected, vindicating our choice of flux--tube overlap in eqn.
\ref{3p0mesondec}.

\section{{\sc \bf Appendix} : Decay amplitudes}

\subsection{Lowest radial states \label{3p0groundsection} }

The lowest radial state wave functions were listed in eqn. \ref{3p0ground}.
Defining

\beqn
\label{3p0amplitude}
\tilde{A} = (\frac{a \tilde{c}}{9 \sqrt{3}} \frac{1}{2} A^{0}_{00}
\sqrt{\frac{fb}{\pi}})
 \; \frac{1}{{(1 + f b / (2 \beta^{2}) \, )}}
 \; 8 \pi^{3/4} \frac{\beta_{A}^{3/2}}{{(2 \beta_{A}^{2}+\beta^{2})}^{5/2}}
\exp (-{(\frac{M}{m+M})}^{2} \frac{p_{B}^{2}}{2 \beta_{A}^{2}+\beta^{2} })
\eeqn
we introduce the partial wave S, P, D and F definitions

\beqna
\label{3p0partial}
P_{1} & = & i 2 \tilde{A} \;  \frac{M}{m+M} p_{B} \;
(\beta^{2} + \frac{m}{M} (2 \beta_{A}^{2}+\beta^{2}) ) \nonumber \\
S_{1} & = & \sqrt{2} \tilde{A} \frac{\beta}{2 \beta_{A}^{2}+\beta^{2}}
 \; (6 \beta_{A}^{2} (2 \beta_{A}^{2}+\beta^{2}) +
2 {(\frac{M}{m+M} p_{B})}^{2} (\beta^{2} + \frac{m}{M} (2
\beta_{A}^{2}+\beta^{2}) ) )  \nonumber \\
D_{1} & = & \sqrt{2} \tilde{A} \frac{\beta}{2 \beta_{A}^{2}+\beta^{2}}
 \; 2 {(\frac{M}{m+M} p_{B})}^{2} \;  (\beta^{2} + \frac{m}{M} (2
\beta_{A}^{2}+\beta^{2}) ) \nonumber \\
S_{2} & = & 2\sqrt{2} \tilde{A} \frac{\beta_{A}}{2 \beta_{A}^{2}+\beta^{2}}
 \; (-3 \beta^{2} (2 \beta_{A}^{2}+\beta^{2}) +
2 {(\frac{M}{m+M} p_{B})}^{2} (\beta^{2} + \frac{m}{M}
(2 \beta_{A}^{2}+\beta^{2}) ) ) \nonumber \\
D_{2} & = & 2\sqrt{2} \tilde{A} \frac{\beta_{A}}{2 \beta_{A}^{2}+\beta^{2}}
 \; 2 {(\frac{M}{m+M} p_{B})}^{2} \;  (\beta^{2} + \frac{m}{M}
(2 \beta_{A}^{2}+\beta^{2}) )  \nonumber \\
P_{2} & = & - i \frac{8}{\sqrt{15}} \tilde{A}
\frac{\beta_{A}^{2}}{{(2 \beta_{A}^{2}+\beta^{2})}^{2}}
 \; \frac{M}{m+M} p_{B} \;  (-5 \beta^{2} (2 \beta_{A}^{2}+\beta^{2}) +
2 {(\frac{M}{m+M} p_{B})}^{2} (\beta^{2} +
\frac{m}{M} (2 \beta_{A}^{2}+\beta^{2}) ) ) \nonumber \\
F_{2} & = & -i \frac{8}{\sqrt{15}} \tilde{A}
\frac{\beta_{A}^{2}}{{(2 \beta_{A}^{2}+\beta^{2})}^{2}}
 \; 2 {(\frac{M}{m+M} p_{B})}^{3} \;  (\beta^{2} +
\frac{m}{M} (2 \beta_{A}^{2}+\beta^{2}) )
\eeqna
in terms of which we list the amplitudes in Table
\ref{3p0amplitudetable}, consistent with the literature \cite{kokoski87}.

\btablec
\caption{Partial wave amplitudes $M_{L} (A \rightarrow BC)$
indicated in terms of the
functions defined in eqn. \protect\ref{3p0partial}
and named in accordance with partial waves {\it L = S, P, D} or {\it F}.
Flavour can be incorporated for various final states by
multiplying $M_{L} (A \rightarrow BC)$  by 1 (for $\D \D, \Ds \Ds$),
$\protect\sqrt{2}$ (for
$DD, \ds \ds, \Ds \D$) or 2 (for $\ds D$).
}
\label{3p0amplitudetable}
\begin{tabular}{|c|c||c|c|}
\hline 
&  $M_{L} (A \rightarrow BC)$  & &  $M_{L} (A \rightarrow BC)$   \\
\hline \hline 
$^{3}S_{1}\rightarrow \; ^{1}S_{0} \; ^{1}S_{0}$ & $P_{1}/ \sqrt{2}$&
$^{3}S_{1}\rightarrow \; ^{3}P_{2} \; ^{1}S_{0}$ & $- D_{1}  / \sqrt{2} $\\

$^{3}S_{1}\rightarrow \; ^{3}S_{1} \; ^{1}S_{0}$ & $      P_{1}$&
$^{3}S_{1}\rightarrow \; ^{3}P_{1} \; ^{1}S_{0}$ & $   S_{1}/\sqrt{3}$\\

$^{3}S_{1}\rightarrow \; ^{3}S_{1} \; ^{3}S_{1}$ & $\sqrt{7}P_{1}/\sqrt{2}$&
                                                 & $D_{1}/{\sqrt{6}}
$\\ \cline{1-2}

$^{3}P_{2}\rightarrow \; ^{1}S_{0} \; ^{1}S_{0}$ & $ D_{2}/\sqrt{3}    $&
$^{3}S_{1}\rightarrow \; ^{1}P_{1} \; ^{1}S_{0}$ & $  S_{1}/{\sqrt{6}}  $\\

$^{3}P_{2}\rightarrow \; ^{3}S_{1} \; ^{1}S_{0}$ & $ D_{2}/\sqrt{2}      $&
                                                 & $-D_{1}/{\sqrt{3}}

$\\ \cline{3-4}

$^{3}P_{1}\rightarrow \; ^{3}S_{1} \; ^{1}S_{0}$ & $ {S_{2}}/{\sqrt{3}}  $&
$^{3}D_{1}\rightarrow \; ^{1}S_{0} \; ^{1}S_{0}$ & $-P_{2}              $\\

                                                  & $ {D_{2}}/{\sqrt{6}}  $&
$^{3}D_{1}\rightarrow \; ^{3}S_{1} \; ^{1}S_{0}$ & ${P_{2}}/{\sqrt{2}}

$\\

$^{3}P_{0}\rightarrow \; ^{1}S_{0} \; ^{1}S_{0}$ & $-{S_{2}}/{\sqrt{6}}  $&
$^{3}D_{1}\rightarrow \; ^{3}S_{1} \; ^{1}S_{0}$ &
$\sqrt{2}P_{2}/\sqrt{5}
$\\

$^{1}P_{1}\rightarrow \; ^{3}S_{1} \; ^{1}S_{0}$ & $ {S_{2}}/{\sqrt{6}}   $&
                                                  &
$3\sqrt{2}F_{2}/\sqrt{5}
$\\

                                                  & $-{D_{2}}/{\sqrt{3}}  $&
& \\
\hline 
\end{tabular}
\etablec

For $\dss$ final states we employ a $^{3}P_{1} / \, ^{1}P_{1}$ mixing
angle $\tilde{\theta}$ according to the convention \cite{isgur85} :

\beqn
\label{3p0tranform}
\left( \barr{c} 1^{+L} \\ 1^{+H} \earr \right) =
\left( \barr{rr} \cos \tilde{\theta} & \sin \tilde{\theta} \\ -\sin
\tilde{\theta} & \cos \tilde{\theta} \earr \right)
\left( \barr{c} ^{1}P_{1} \\ ^{3}P_{1} \earr \right)
\eeqn
where $1^{+L}$ and $1^{+H}$ indicate the low and high mass states
respectively.

\subsection{Radial excitations}

Instead of deriving the decay amplitudes for radially excited mesons
afresh, it is more immediate to notice that the
radially excited wave functions can be related to the lowest radial
wave functions by differentiation \cite{barbieri75} :

\[ \psi_{2S}(\br) = \frac{1}{\sqrt{3!}} \frac{\beta^{3/2}}{\pi^{3/4}}
\: \{ 3 - 2 \beta^{2} r^{2} \} \: e^{-\frac{1}{2} \beta^{2} r^{2}}
=  \frac{4}{\sqrt{3!}} \beta^{2} \frac{d}{d \beta^{2}} \psi_{S}(\br)
\]
\intereqnvspace

\[ \psi_{3S}(\br) = \frac{1}{\sqrt{5!}} \frac{\beta^{3/2}}{\pi^{3/4}}
\: \{ 15 - 20 \beta^{2} r^{2} + 4 \beta^{4} r^{4} \} \:
e^{-\frac{1}{2}
\beta^{2} r^{2}}
=  \frac{2}{\sqrt{5!}} \: \{3 + 8 \beta^{2} \frac{d}{d \beta^{2}}
   + 8 \beta^{4} \frac{d^{2}}{d^{2} \beta^{2}} \} \: \psi_{S}(\br) \]
\intereqnvspace

\[ \psi_{2P}(\br) = 4 \sqrt{\frac{2!}{5!}} \frac{\beta^{5/2}}{\pi^{1/4}}
\, r \: \{ 5 - 2 \beta^{2} r^{2} \} \: Y_{1 M_{L}} (\hat{\br}) \,
e^{-\frac{1}{2} \beta^{2} r^{2}}
=  2 \sqrt{\frac{2}{5}} \beta^{2} \frac{d}{d \beta^{2}} \psi_{1P}(\br)
\] \intereqnvspace

\[ \psi_{2D}(\br) = 8 \sqrt{\frac{3!}{7!}} \frac{\beta^{7/2}}{\pi^{1/4}}
\, r^{2} \: \{ 7 - 2 \beta^{2} r^{2} \} \: Y_{2 M_{L}} (\hat{\br}) \,
e^{-\frac{1}{2} \beta^{2} r^{2}}
=  2 \sqrt{\frac{2}{7}} \beta^{2} \frac{d}{d \beta^{2}} \psi_{1D}(\br)
\] \intereqnvspace

\beqna \label{3p0radial} \psi_{4S}(\br) & = & -\frac{1}{\sqrt{7!}}
\frac{\beta^{3/2}}{\pi^{3/4}}
\: \{ 105 - 210 \beta^{2} r^{2} + 84 \beta^{4} r^{4} - 8 \beta^{6}
r^{6}\} \: e^{-\frac{1}{2} \beta^{2} r^{2}}
\nonumber \\ & = &  -\frac{8}{\sqrt{7!}} \: \{21 \beta^{2}
\frac{d}{d \beta^{2}}
   + 24 \beta^{4} \frac{d^{2}}{d^{2} \beta^{2}}
   + 8 \beta^{6} \frac{d^{3}}{d^{3} \beta^{2}} \} \: \psi_{S}(\br) \eeqna

The differential operator depends only on $\beta$, and can hence be
pulled out when the integration in eqn. \ref{3p0master} over $\br_{A}$
and $\by$ is performed. The amplitudes for radially excited meson
decay can hence be found by applying the differential operators in
eqn. \ref{3p0radial} to the amplitudes in table \ref{3p0amplitudetable}.

\vspace{1cm}
{\noindent \Large \bf Figure Captions \vspace{0.35cm}}

\noindent Figure 1: The contours indicate $\psi(3770)$ total hadronic
widths
fitting the
mean experimental width in eqn. \ref{3p0experiment}.
Where there is 1D-2S mixing with mixing angle $\theta$, we display
$\sin \theta$ (which can be positive or negative). Pure 1D and 2S
states
are also shown.
The ``square'' indicates theoretically acceptable parameters.

\vspace{0.3cm}

\noindent Figure 2: Total hadronic widths of the higher mass charmonia
$\psi(4040)$, $\psi(4160)$ and $\psi(4415)$ fitting experiment. For
the latter two states the contours enclose narrow regions with
experimentally acceptable
total widths, as indicated by shading in one case. They
indicate 1$\sigma$ variations from
the mean experimental width (eqn. \ref{3p0experiment}). We display
$\psi(4160)$ as 3S or 2D, and $\psi(4415)$ as 4S. The regions in table
\ref{3p04040table} where $\psi(4040)$  fits
experiment are indicated by three large ``dots''.  The ``square''
indicates theoretically acceptable parameters.

\vspace{0.3cm}

\noindent Figure 3: Total hadronic widths of the 2P states with
$J^{PC} = 2^{++}$
(3.98 GeV), $1^{++}$ (3.95 GeV) and $0^{++}$
(3.92 GeV) \cite{isgur85} for theoretically acceptable
parameters. We indicate regions where these states may be narrow. The
lighter and darker shading indicate $2^{++}$ widths less than 5 MeV
and 1 MeV respectively. The contours for $0^{++}$ and $1^{++}$ enclose
narrow regions where widths are less than 5 MeV or 1 MeV.

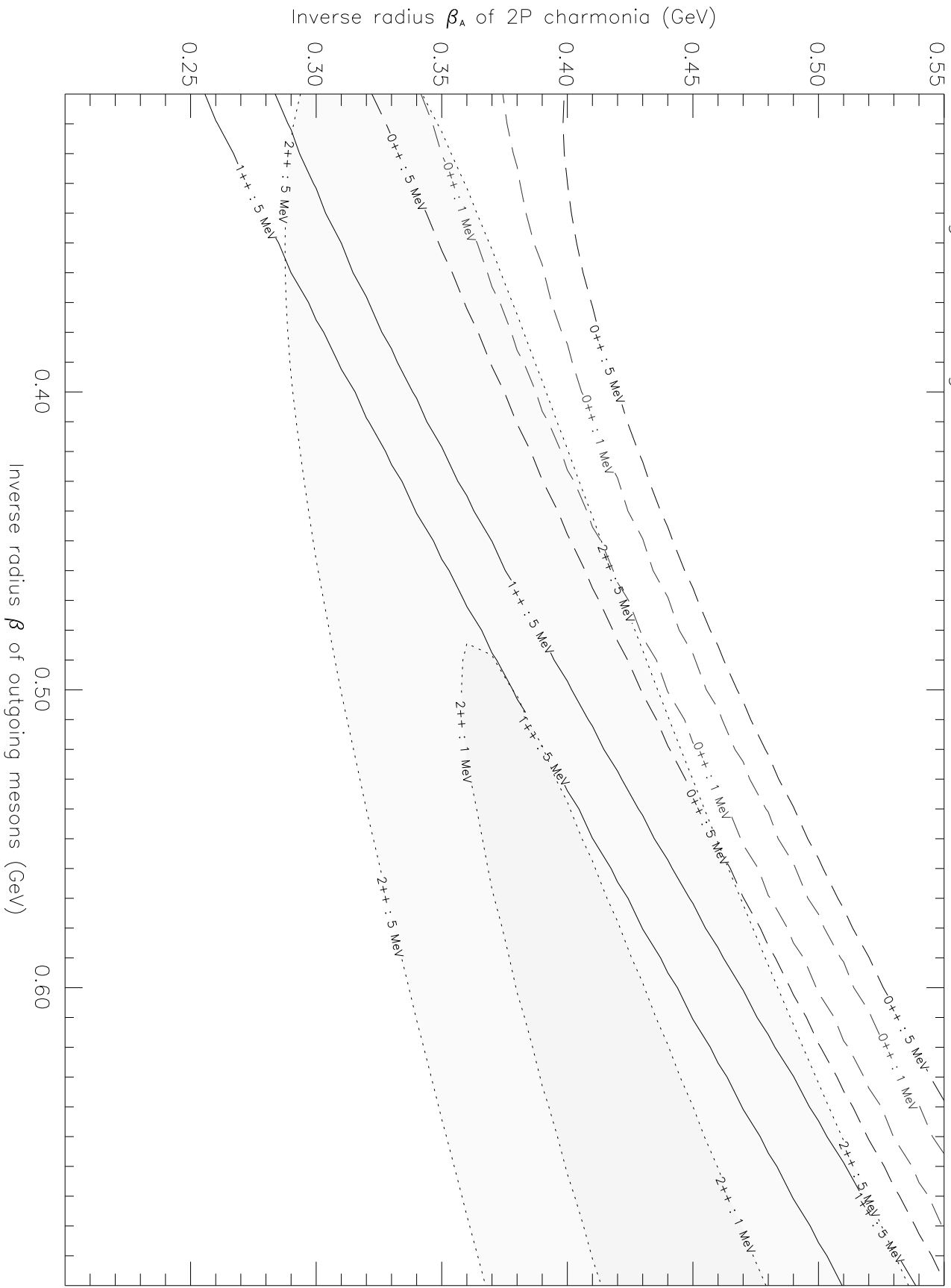
\end{document}